\documentclass[11pt,a4paper]{article}
\usepackage[hyperref]{acl2020}
\usepackage{times}
\usepackage{latexsym}
\usepackage{graphicx}

\usepackage{microtype}

\title{MSV Challenge 2022: NPU-HC Speaker Verification System for Low-resource Indian Languages}

\author{Yue Li$^1$, Li Zhang$^1$, Namin Wang$^2$, Jie Liu$^2$, Lei Xie$^{1}$\thanks{* Corresponding author.} \\
        $^1$Audio, Speech and Language Processing Group (ASLP@NPU), School of Computer Science, \\
        Northwestern Polytechnical University (NPU), China \\ 
        $^2$Huawei Cloud \\
        \texttt{riollee.yli@gmail.com}, ~ \texttt{lxie@nwpu.edu.cn}
}

\date{}

\begin{document}
\maketitle
\begin{abstract}

This report describes the NPU-HC speaker verification system submitted to the  O-COCOSDA Multi-lingual Speaker Verification (MSV) Challenge 2022, which focuses on developing speaker verification systems for low-resource Asian languages. 
We participate in the I-MSV track, which aims to develop speaker verification systems for various Indian languages. 
In this challenge, we first explore different neural network frameworks for low-resource speaker verification. 
Then we leverage vanilla fine-tuning and weight transfer fine-tuning to transfer the out-domain pre-trained models to the in-domain Indian dataset. 
Specifically, the weight transfer fine-tuning aims to constrain the distance of the weights between the pre-trained model and the fine-tuned model, which takes advantage of the previously acquired discriminative ability from the large-scale out-domain datasets and avoids catastrophic forgetting and overfitting at the same time. 
Finally, score fusion is adopted to further improve performance.
Together with the above contributions, we obtain 0.223\% EER on the public evaluation set, ranking 2nd place on the leaderboard.
On the private evaluation set, the EER of our submitted system is 2.123\% and 0.630\% for the constrained and unconstrained sub-tasks of the I-MSV track, leading to the 1st and 3rd place in the ranking, respectively.

\end{abstract}


\section{Introduction}

Speaker verification (SV) is the task of verifying whether an input utterance matches the claimed identity \cite{rosenberg1976automatic}. 
In recent years, deep learning has achieved remarkable success in SV tasks, but current methods usually rely on a huge amount of labeled training data \cite{nagrani2017voxceleb,fan2020cn}. 
However, obtaining massive labeled data for every language is a time-consuming and costly task.
Therefore, the 2022 MSV challenge has been particularly designed for understanding and comparing current SV techniques in low-resource Asian languages, where the labeled speaker data is limited in quantity.
Specifically, the challenge includes two evaluation tracks, A-MSV and I-MSV.
The former focuses on the development of SV systems in various Asian languages while the latter particularly focuses on SV for Indian languages.
In this challenge, we participate in the I-MSV track, which includes the constrained and unconstrained  sub-tasks. For the constrained task, the speaker verification model can be trained only with the given fixed training set given by the challenge organizer, while extra training data can be used for the unconstrained task.
In particular, the fixed training set consists of 1000 audio recordings spoken by 100 speakers, about 155 hours in total, which is not enough to train a robust speaker verification system from scratch. 
So the major challenge in the I-MSV track is the data limitation.

Low data resource speaker verification has drawn much attention recently. 
The straightforward idea for the low resource problem is to leverage high resource labeled datasets (e.g., from another language) to pre-train the speaker verification models ~\cite{zhang2020npu, gusev2020stc, shahnawazuddin2021children}. 
However, bringing in additional training datasets usually leads to a domain mismatch problem, which means that there is a distribution change or domain shift between two domains that degrades the performance of SV systems \cite{wang2018deep}. 
To deal with domain mismatch problems, recent approaches include domain adversarial training \cite{wang2018unsupervised,rohdin2019speaker}, back-end processing \cite{garcia2014unsupervised, lee2019coral+}, and fine-tuning strategy \cite{zhang2021multi,tong2020jd}.

In this challenge, we first explore different popular speaker verification models in the constrained I-MSV sub-task, most of which are variants of ECAPA-TDNN \cite{desplanques2020ecapa} and Resnet34 \cite{heo2020clova}. Particularly for the unconstrained I-MSV, we first leverage high resource English language datasets, VoxCeleb1\&2 \cite{nagrani2017voxceleb, nagrani2020voxceleb}, as pre-trained datasets to improve the performance of SV systems.
Then to address the domain mismatch problem induced by the language difference, we study vanilla fine-tuning and weight transfer fine-tuning ~\cite{zhangnpu, zhangnpu1} strategies to transfer the pre-trained model to the in-domain model with the given Indian language dataset. Specifically, the weight transfer fine-tuning~\cite{zhangnpu, zhangnpu1} aims to constrain the distance of the weights between the pre-trained model and the fine-tuned model to mitigate catastrophic forgetting and overfitting problems during fine-tuning. In addition, we also use score average fusion to improve the performance of our SV systems. The experimental results demonstrate the effectiveness of the above methods and we finally get 1st and 3rd place in the constrained and the unconstrained I-MSV sub-tasks, respectively. Our final code \footnote{https://github.com/RioLLee/MSVChallenge} is based on the SpeechBrain \cite{ravanelli2021speechbrain}.

\section{Methodology}

In this section, we describe different neural network frameworks in low-resource speaker verification and the fine-tuning methods for domain adaptation from out-domain pre-trained models to the in-domain Indian dataset.
Moreover, we introduce weight transfer fine-tuning, which constrains the distance between the weights of the pre-trained model and the fine-tuned model, to improve the performance of the speaker verification systems.

\subsection{ECAPA-TDNN}

ECAPA-TDNN is known as one of the state-of-the-art speaker embedding models. 
This model has achieved striking performance in many speaker verification challenges with a large amount of labeled datasets\cite{zhang2021beijing,desplanques2020ecapa,thienpondt2021idlab}.
Therefore, in the 2022 I-MSV challenge, we explore two kinds of ECAPA-TDNN with different channels of 1024 and 2048 in low-resource speaker verification.

As shown in Table \ref{tab:backbone_ecapa}, ECAPA-TDNN consists of three SE-Res2Block layers, a multi-layer aggregation layer, and a channel- and context-dependent statistics pooling layer\cite{desplanques2020ecapa}. In Table \ref{tab:backbone_ecapa}, $ T $ refers to the length of the input feature, while $ C $ is the channels of the convolution neural network. $ D $ is the embedding dimension. The loss function we use in this report is additive angular margin softmax (AAMSoftmax) loss \cite{deng2019arcface} and the N is the speaker numbers of training datasets.






\begin{table}[th]
\centering
\footnotesize
\caption{ECAPA-TDNN Structure}
\label{tab:backbone_ecapa}
\resizebox{\linewidth}{!}{
\begin{tabular}{lclclcl}
\hline
Layer    & \multicolumn{2}{c}{Kernel Size}               & \multicolumn{2}{c}{Stride}         & \multicolumn{2}{c}{Output Shape}                       \\ \hline
Conv1D    & \multicolumn{2}{c}{$ 5 $}   & \multicolumn{2}{c}{$ 1 $} & \multicolumn{2}{c}{$ T \times C $}          \\ \hline
SE-Res2Block1     & \multicolumn{2}{c}{$ 3 $}   & \multicolumn{2}{c}{$ 2 $} & \multicolumn{2}{c}{$ T \times C $}          \\ \hline
SE-Res2Block2     & \multicolumn{2}{c}{$ 3 $}   & \multicolumn{2}{c}{$ 3 $} & \multicolumn{2}{c}{$ T \times C $}          \\ \hline
SE-Res2Block3     & \multicolumn{2}{c}{$ 3 $}   & \multicolumn{2}{c}{$ 4 $} & \multicolumn{2}{c}{$ T \times C $}          \\ \hline
Conv1D     & \multicolumn{2}{c}{$ 1 $}  & \multicolumn{2}{c}{$ 1 $} & \multicolumn{2}{c}{$ T \times (3 \times C) $} \\ \hline
ASP      & \multicolumn{2}{c}{-}                         & \multicolumn{2}{c}{-}              & \multicolumn{2}{c}{ ($6 \times C) $}                               \\ \hline
Linear   & \multicolumn{2}{c}{1}                       & \multicolumn{2}{c}{-}              & \multicolumn{2}{c}{ $ D $}                                \\ \hline
AAMSoftmax   & \multicolumn{2}{c}{-}                       & \multicolumn{2}{c}{-}              & \multicolumn{2}{c}{$N$}                                \\ \hline
\end{tabular}
}
\vspace{-0.5cm}
\end{table}

\subsection{ResNet34-SE}

The deep residual network (ResNet) is a well-known deep neural network that solves the problem of gradient disappearance with short-cut connections. 
In recent years, ResNet becomes a popular backbone in the speaker verification field \cite{heo2020clova,zhang2021beijing}. 
In this challenge, we try two kinds of ResNet models with squeeze-and-excitation (SE) \cite{hu2018squeeze} attention and a variant of SE attention \cite{zhang2021duality}. 
The model structure of the ResNet34-SE is illustrated in Table \ref{tab:backbone_resnet}. 
Specially, we use the attentive statistics pooling (ASP) \cite{okabe18_interspeech} as the pooling layer in ResNet34-SE.




\begin{table}[th]
\centering
\footnotesize
\caption{ResNet34-SE Structure}
\label{tab:backbone_resnet}
\resizebox{\linewidth}{!}{
\begin{tabular}{lclclcl}
\hline
Layer    & \multicolumn{2}{c}{Kernel Size}               & \multicolumn{2}{c}{Stride}         & \multicolumn{2}{c}{Output Shape}                       \\ \hline
Conv2D    & \multicolumn{2}{c}{$ 3\times 3 $}   & \multicolumn{2}{c}{$ 1 \times 1 $} & \multicolumn{2}{c}{$ T \times 80 \times C $}          \\ \hline
Res1     & \multicolumn{2}{c}{$ 3 \times 3 $}   & \multicolumn{2}{c}{$ 1 \times 1 $} & \multicolumn{2}{c}{$ T \times 80 \times C $}          \\ \hline
SE-Module & \multicolumn{2}{c}{-}                         & \multicolumn{2}{c}{-}              & \multicolumn{2}{c}{$ T \times 80 \times C $}          \\ \hline
Res2     & \multicolumn{2}{c}{$ 3 \times 3 $}  & \multicolumn{2}{c}{$ 2 \times 2 $} & \multicolumn{2}{c}{$ T \times 40 \times C $}          \\ \hline
SE-Module & \multicolumn{2}{c}{-}                         & \multicolumn{2}{c}{-}              & \multicolumn{2}{c}{$ T \times 40 \times C $}          \\ \hline
Res3     & \multicolumn{2}{c}{$ 3 \times 3 $}  & \multicolumn{2}{c}{$ 2 \times 2 $} & \multicolumn{2}{c}{$ T_{/2} \times 20 \times C $} \\ \hline
SE-Module & \multicolumn{2}{c}{-}                         & \multicolumn{2}{c}{-}              & \multicolumn{2}{c}{${ T_{/2} \times 20 \times C}$} \\ \hline
Res4     & \multicolumn{2}{c}{$ 3 \times 3  $} & \multicolumn{2}{c}{$ 2 \times 2 $} & \multicolumn{2}{c}{${T_{/4} \times 10 \times C}$}  \\ \hline
SE-Module & \multicolumn{2}{c}{-}                         & \multicolumn{2}{c}{-}              & \multicolumn{2}{c}{${T_{/4} \times 10 \times C}$}  \\ \hline
Flatten  & \multicolumn{2}{c}{-}                         & \multicolumn{2}{c}{-}              & \multicolumn{2}{c}{${T_{/8} \times ( 10 \times C)}$}           \\ \hline
ASP      & \multicolumn{2}{c}{-}                         & \multicolumn{2}{c}{-}              & \multicolumn{2}{c}{$( 10 \times C)$}                               \\ \hline
Linear   & \multicolumn{2}{c}{1}                       & \multicolumn{2}{c}{-}              & \multicolumn{2}{c}{$D$}                                \\ \hline
AAM-Softmax   & \multicolumn{2}{c}{-}                       & \multicolumn{2}{c}{-}              & \multicolumn{2}{c}{$ N $}                                \\ \hline
\end{tabular}
}
\vspace{-0.5cm}
\end{table}


\subsection{Fine-tuning}

Leveraging additional out-domain datasets leads to domain mismatch problems between the large out-domain datasets and the small in-domain Indian dataset ~\cite{qin2021our, zhangnpu, zhangnpu1}. The mismatch lies in various aspects including cross-language differences and cross-recording-device differences.
Vanilla fine-tuning is the most common approach to deal with domain mismatch.  The process of fine-tuning is to initialize the weights of the model to be fine-tuned with those of the pre-trained model and then train this model with the target-domain dataset.
Specifically, in the unconstrained I-MSV, we use the VoxCeleb1\&2 development sets to pre-train the speaker verification models and then fine-tune the models with the Indian language training set to drag the models to the target domain as well as to maintain their discrimination ability.

However, vanilla fine-tuning just initializes the weights of the fine-tuned model with those of the pre-trained model without considering the catastrophic forgetting
and overfitting problems. Therefore, we introduce a weight transfer loss as in~\cite{zhangnpu, zhangnpu1} to deal with the above problems, which constrains the distance between the weights of the pre-trained model and those of the fine-tuned model during the fine-tuning process.
Specifically, suppose the weights of the pre-trained model and fine-tuning model are $W^s$ and $W^t$ respectively, the weight transfer loss $L_{wt}$ is calculated as
\begin{equation}
    L_{w t}=\left\|W^{s}-W^{t}\right\|_{2}
\end{equation}
Finally, the final loss function during fine-tuning is
\begin{equation}
   L_{ft} = L_{CE} +   L_{wt} + L_2,
\label{eqs1}
\end{equation}
where $L_{CE}$ is the speaker classification loss~(AAMSoftmax) and $L_{2}$ is the common L2 regularization loss.



\section{Experiments \& Analysis}

\subsection{Datasets \& Augmentation}

In the I-MSV track, the development data consists of speech data in Indian languages, collected in multiple sessions using five different sensors. In the evaluation set, the enrolment data consists of utterances from the English language captured in multiple sessions using only a headphone as the sensor. There are two test sets, which are the public test dataset and the private test dataset, provided with language and recording device mismatch compared with the enrollment dataset.

In the constrained sub-task of the I-MSV track, we only use the released development Indian dataset as the training set for our speaker verification models. 
In the unconstrained sub-task of the I-MSV track, we leverage the VoxCeleb1\&2 \cite{nagrani2017voxceleb, nagrani2020voxceleb} as our pre-trained datasets.
Then we fine-tune the pre-trained models with the released Indian dataset.

Online data augmentation \cite{cai2020fly} is used for all our speaker verification models. Specifically, we adopt frequency-domain specAug \cite{park2019specaugment}, time warping specAug, additive noise augmentation \cite{snyder2015musan}, and reverberation augmentation \cite{habets2006room}. The details of the augmentation configurations are listed as follows:

\begin{itemize}
    \item Frequency-Domain SpecAug: We apply time and frequency masking as well as time warping to the input spectrum (frequency-domain implementation) \cite{park2019specaugment}.
    \item Additive Noise: We add the noise, music, and babble types from MUSAN \cite{snyder2015musan} to the original speech.
    \item Reverberation: We simulate reverberant speech by convolving clean speech with different RIRs from \cite{habets2006room}.
    \item Speed perturb: We adopt speed perturbation (0.9 and 1.1 times) in the training stage.
\end{itemize}

\subsection{Experimental Setup}

\textbf{Model Configuration}
The channel numbers of TDNN layers in ECAPA-TDNN are 1024 or 2048, and the dimensions of the embedding layer are 192 or 256 respectively.
For the ResNet34, we train 5 ResNet34-related models with a similar structure, which have \{64, 128, 256, 512\} or \{32, 64, 128, 256\} channels of residual blocks and multi-head attention statistic pooling. 
In particular, we added SE blocks with 8 reductions to the last layer of the residual block in our models.

\noindent \textbf{Training Details} Eighty-dimensional Mel-filter bank features with 25ms window size and 10ms window shift are extracted as model inputs. 
During the training stage, the learning rate of all models training varies between 1e-8 and 1e-3 using the triangular2 policy \cite{smith2017cyclical} and the optimizer is Adam \cite{kingma2014adam}. 
The hyperparameter scale and margin of AAM-softmax are set to 30 and 0.2 respectively. To prevent overfitting, we apply a weight decay of 2e-4 on all weights in the models.

\noindent \textbf{Score Average Fusion} We split enroll audio recordings by a random length between 10 and 60 seconds, and then we average the speaker embeddings extracted from all audio recordings of the same speaker as the embedding of this speaker. 
To further improve the performance of the speaker verification systems, we use score average fusion based on the performance of the models on the public test set.

\noindent \textbf{Score Metric}
In the test phase, we use cosine similarity as the scoring criterion. The performance metric is equal error rate (EER) \cite{reynolds20172016}.

\subsection{Experimental Results}

We evaluate all models with the above-mentioned strategies on the public test datasets of the constrained I-MSV and the unconstrained I-MSV. The results are summarized in Table \ref{tab:constrained_result} and Table \ref{tab:unconstrained_result} respectively. 
For the results of the constrained I-MSV, as shown in Table \ref{tab:constrained_result}, the best single model is ECAPA\_2048 with the lowest EER of 1.764\% among all speaker verification models. 
After fusing the scores from ECAPA\_1024, ECAPA\_2048, and ResNet34SE\_256, we obtain the best fusion EER of 1.677\%.
The score average fusion model gets a relative EER reduction by 4\% compared with the best single model.

\begin{table}[htbp] 
    \caption{\emph{EER of all systems on the public test dataset of the constrained I-MSV}}
    \small
    \renewcommand{\arraystretch}{1.2}
    \setlength\tabcolsep{15.3pt}
    \begin{tabular}{lll}
        \hline
        Index & Model            & EER(\%)     \\ 
        \hline
        A            & ECAPA\_1024           & 1.881   \\
        B            & ECAPA\_2048           & \textbf{1.764}    \\
        C            & ResNet34SE\_512       & 2.030    \\
        D            & ResNet34SE\_256       & 1.864    \\
        E            & ResNetDTCF\_512            & 1.899  \\ 
        \hline
        Fusion       & {[}A+B+D{]}           & \textbf{1.677} \\ 
        \hline
    \end{tabular}
    \label{tab:constrained_result}
\end{table}

In Table~\ref{tab:unconstrained_result}, the ResNet34SE\_512\_fine-tune model achieves the best single model result with EER of 0.29\%. 
Finally, fusing scores from ECAPA\_2048\_weight\_transfer and ResNet34SE\_512\_fine-tune leads to our best EER of 0.223\%, which achieves a relative EER reduction of 23\% compared with the single best model. 
On the other hand, we can find that the performance of the fine-tuned models improves a lot compared to that of the models trained from scratch. 
The most superior model is ResNet34SE\_512 with a dramatic drop in EER, where the EER of the unconstrained I-MSV is relatively 86\% lower than that of the constrained I-MSV.

\begin{table}[htbp] 
    \caption{\emph{EER of all systems on the public test dataset of the unconstrained I-MSV}}
    \renewcommand{\arraystretch}{1.2}
    \resizebox{\linewidth}{!}{
        \begin{tabular}{lll}
            \hline
            Index & Model                & EER(\%)      \\ 
            \hline
            A1          & ECAPA\_1024\_fine-tune      & 0.510    \\
            A2          & ECAPA\_1024\_weight\_transfer            & 0.508    \\
            B1          & ECAPA\_2048\_fine-tune      & 0.680    \\
            B2          & ECAPA\_2048\_weight\_transfer            & 0.324    \\
            C1          & ResNet34SE\_512\_fine-tune  & \textbf{0.289}    \\
            C2          & ResNet34SE\_512\_weight\_transfer            & 0.494    \\
            D1          & ResNet34SE\_256\_fine-tune  & 0.693    \\
            D2          & ResNet34SE\_256\_weight\_transfer            & 0.862    \\
            E1          & ResNetDTCF\_512\_fine-tune      & 1.216    \\ 
            E2          & ResNetDTCF\_512\_weight\_transfer            & 0.561    \\
            \hline
            Fusion      & {[}B2+C1{]}             & \textbf{0.223}   \\ 
            \hline
        \end{tabular}
    }
    \label{tab:unconstrained_result}
\end{table}

As shown in Table \ref{tab:private_result}, our best-submitted model is the fused model for the private test of the constrained I-MSV, which achieves the EER of 2.123\%. 
For the unconstrained I-MSV, the best model is ResNet34SE\_512\_fine-tune with the EER of 0.630\%.

\begin{table}[htbp] 
    \caption{\emph{EER of submitted systems on the private test dataset}}
    \renewcommand{\arraystretch}{1.2}
    \resizebox{\linewidth}{!}{
        \begin{tabular}{lll} 
            \hline
            Track & Model & EER(\%)      \\ 
            \hline
            Constrained I-MSV   & fused model       & 2.123    \\
            Unconstrained I-MSV  & ResNet34SE\_512\_fine-tune          & 0.630    \\
            \hline
        \end{tabular}
    }
    \label{tab:private_result}
\end{table}

\section{Discussion}

This paper introduces the main approaches used in our submitted systems for the MSV challenge 2022, especially exploring the effectiveness of ECAPA-TDNN and ResNet34-SE models in SV for low resource Indian languages, vanilla fine-tuning and weight transfer fine-tuning strategies to transfer pre-trained models into the Indian dataset as well as score average fusion.
Through our study, we can find that there is still substantial space for improving the performance of speaker verification for low resource languages. 
For instance, for the constrained I-MSV, we plan to explore the recent low-resource learning strategies, such as few-shot learning \cite{wang2020generalizing, yang2022domain}.
Moreover, for the unconstrained I-MSV, it is a promising method to use data augmentation strategies such as cross-lingual voice conversion \cite{shahnawazuddin2020voice} to expand the data size.  

\section{Conclusion}

In this report, we describe our submission for the I-MSV track of the 2022 Multilingual Speaker Verification (MSV) Challenge.
In this challenge, we first explore ECAPA-TDNN and ResNet34-SE in the low-resource Indian language speaker verification, with the conclusion that all of these models outperform the baseline model.
Moreover, vanilla fine-tuning and weight transfer fine-tuning are introduced to deal with the domain mismatch between the in-domain Indian dataset and the large-scale out-domain datasets.
Finally, score fusion is beneficial to our speaker verification systems developed for the Indian languages according to the experiments.
Together with the above approaches, our final EER of the constrained/unconstrained I-MSV achieves 2.123\%/0.630\% and we finally take the 1st and the 3rd place in the rankings in the constrained and the unconstrained tasks respectively.

\bibliography{acl2020}
\bibliographystyle{acl_natbib}

\appendix

\end{document}